\documentclass[a4paper,
               biblatex,     % biblatex is used
               keeplastbox,   % flushend option: not to un-indent last line in References
               ]{jacow}
\usepackage{pdfpages,multirow,ragged2e} %
\makeatletter%
	\ifboolexpr{bool{xetex}}
	 {\renewcommand{\Gin@extensions}{.pdf,%
	                    .png,.jpg,.bmp,.pict,.tif,.psd,.mac,.sga,.tga,.gif,%
	                    .eps,.ps,%
	                    }}{}
\makeatother

\ifboolexpr{bool{xetex} or bool{luatex}} % test for XeTeX/LuaTeX
 {}                                      % input encoding is utf8 by default
 {\usepackage[utf8]{inputenc}}           % switch to utf8

\usepackage[USenglish]{babel}

\ifboolexpr{bool{jacowbiblatex}}%
 {%
  \addbibresource{MOPC72.bib}
 }{}
\begin{document}

\title{Advancing electron injection dynamics and mitigation approaches in the Electron-Ion 
Collider’s swap-out injection scheme\thanks{Work supported by 
Brookhaven Science Associates, LLC under Contract No. DE-SC0012704 with the U.S. Department of Energy}}

\author{Derong Xu\thanks{dxu@bnl.gov}, Ferdinand Willeke, Michael M. Blaskiewicz, Yun Luo, Christoph Montag \\
Brookhaven National Laboratory, Upton, NY, USA}
	
\maketitle

\begin{abstract}
The Electron-Ion Collider (EIC) will use swap-out injection scheme for the Electron Storage Ring (ESR) 
to overcome limitations in polarization lifetime. However, the pursuit of highest luminosity with the 
required $28~\mathrm{nC}$ electron bunches encounters stability challenges in the Rapid Cycling 
Synchrotron (RCS). One method is to inject multiple RCS bunches into a same ESR bucket.
In this paper we perform simulation studies investigating proton emittance growth and 
electron emittance blowup in this injection scheme.
Mitigation strategies are explored. 
These findings promise enhanced EIC stability and performance, shaping potential future 
operational improvements.
\end{abstract}

\section{Introduction}
The EIC, to be constructed at Brookhaven National Laboratory (BNL), 
is designed to facilitate collisions between polarized high-energy electron 
beams and hadron beams \cite{willeke2021electron}.
The highest luminosity of $10^{34}~\mathrm{cm}^{-2}\mathrm{s}^{-1}$
will be achieved by colliding $10~\mathrm{GeV}$ electrons and $275~\mathrm{GeV}$ protons.
The corresponding beam parameters are shown in Table~\ref{tab:SimulationParameters}.
The physics program calls for the simultaneous storage of electron bunches with both
spin helicities. A full energy polarized electron injector is needed, so that the electron bunches are
injected into the Electron Storage Ring (ESR) with high transverse polarization and the desired
spin direction.

\begin{table}
  \centering
  \caption{Beam parameters for the highest luminosity from EIC-CDR \cite{willeke2021electron}. 
  ``H'' stands for horizontal, ``V'' denotes vertical, and ``L'' indicates longitudinal parameters below.}
  \begin{tabular}{lccc}
  \toprule
    Parameter & Unit & Proton & Electron \\
    \hline
    Circumference & $\mathrm{m}$ & \multicolumn{2}{c}{$3834$}  \\
    Energy & $\mathrm{GeV}$ & $275$ & $10$\\
    Particles per bunch & $10^{11}$ & $0.688$ & $1.72$\\
    Crossing angle & $\mathrm{mrad}$ & \multicolumn{2}{c}{$25.0$}\\
    $\beta_x^*/\beta_y^*$ & $\mathrm{cm}$ & $80.0/7.20$ & $45.0/5.60$\\
    H. RMS emittance & $\mathrm{nm\cdot rad}$ & $11.3$ & $20.0$\\
    V. RMS emittance & $\mathrm{nm\cdot rad}$ & $1.00$ & $1.29$\\
    H. size & $\mathrm{\mu m}$ & \multicolumn{2}{c}{$95.0$}\\
    V. size & $\mathrm{\mu m}$ & \multicolumn{2}{c}{$8.5$}\\
    Bunch length & $\mathrm{cm}$ & $6.0$ & $2.0$\\
    Energy spread & $10^{-4}$ & $6.6$ & $5.5$\\
    H. tune & - & $0.228$ & $0.08$\\
    V. tune & - & $0.210$ & $0.14$\\
    L. tune & - & $-0.010$ & $-0.069$\\
    H. Damping time & turns & - & $4000$\\
    V. Damping time & turns & - & $4000$\\
    L. damping time & turns & - & $2000$\\
  \bottomrule
  \end{tabular}
  \label{tab:SimulationParameters}
\end{table}

The Rapid Cycling Synchrotron (RCS) will serve as the electron accumulator, ramping electrons 
from $400~\mathrm{MeV}$ up to $18~\mathrm{GeV}$. It will be located within the same existing tunnel. 
The RCS features a $96$-fold lattice periodicity design to avoid spin imperfection and 
intrinsic resonances, ensuring the maintenance of electron polarization.

Due to the short lifetime of polarization, the ESR injection scheme adopts full bunch swap,
necessitating that the RCS provides $28~\mathrm{nC}$ per bunch, 
as indicated in Table~\ref{tab:SimulationParameters}. However,
electron bunches with $28~\mathrm{nC}$ exhibit instability at the lower energy of 
$400~\mathrm{MeV}$. There is ongoing consideration regarding the development 
of a dedicated booster to mitigate this issue \cite{lovelace2023damping}. 
Nonetheless, achieving $28~\mathrm{nC}$ per bunch is beyond the state of 
the art for such boosters \cite{calvey2021measurements}.

Another approach involves injecting multiple electron bunches into one same ESR bucket, 
where synchrotron radiation damping will eventually merge the injected and stored bunches. 
According to Liouville's theorem, the injected bunch cannot occupy the same phase-space volume as 
the stored beam without impacting the latter. Therefore, a separation between the injected and 
stored beams is necessary. Previous study has demonstrated that proton emittance growth occurs 
when injection errors are included \cite{qiang:ipac2021-wepab252}. Additionally, the electron emittance is likely to blow up 
significantly due to the large tune spread resulting from the electromagnetic kick from the 
opposing proton beam.

In this paper, we will study the feasibility of this injection scheme in presence of beam-beam
interaction. The beam parameters used in our simulation is 
presented in Tab.~\ref{tab:SimulationParameters}.

\begin{figure*}[!htbp]
    \centering
    \includegraphics[width=0.9\textwidth]{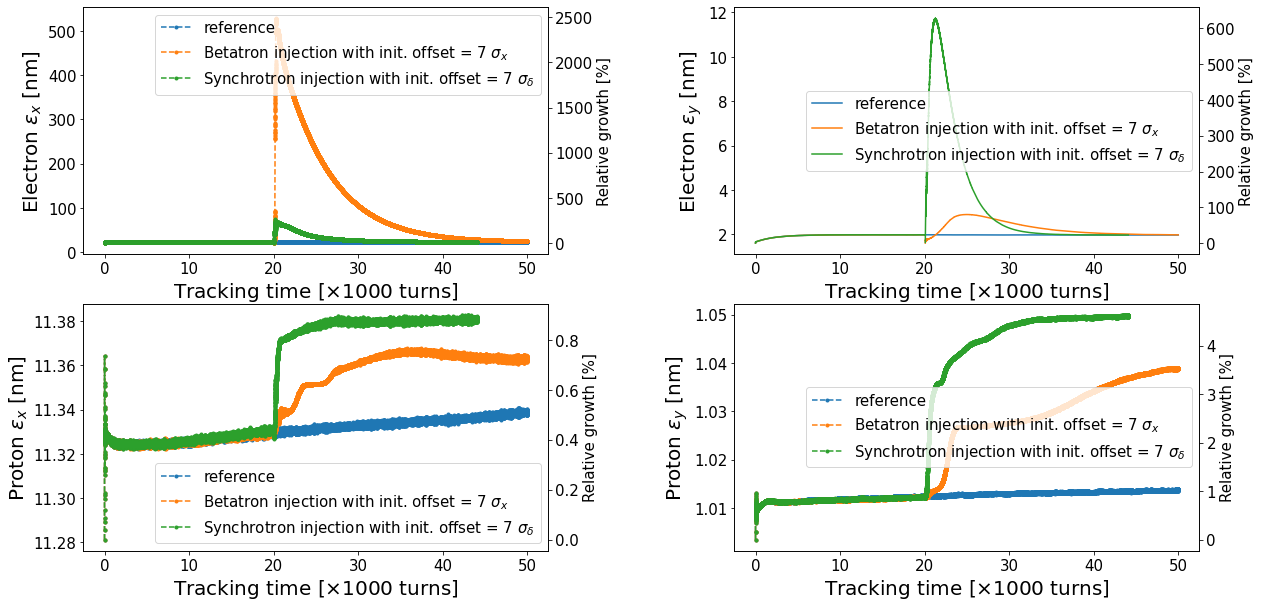}
    \caption{Emittance evolution in strong-strong simulation for betatron 
    and synchrotron injection.}
    \label{fig:ss}
\end{figure*}

\section{Betatron vs synchrotron injection}
According to the methods at the injection point to separate the injected and 
stored bunches, we adopt the terms ``betatron injection'' 
and ``synchrotron injection'' as described in \cite{mori2012design}.

In betatron injection, the injected bunch is positioned at the same location in the 
longitudinal phase space and undergoes betatron oscillation due to its initial 
transverse offset from the design orbit. Conversely, in synchrotron injection, 
the injected bunch differs in energy from the stored beam, with the two beams 
separated by a dispersion at the injection point. 
This leads to the centroid of the injected bunch performing synchrotron oscillation.

Figure~\ref{fig:ss} displays the strong-strong simulation results for betatron and synchrotron injection. 
The old electron bunch is tracked for $20,000$ turns before being kicked out. Subsequently, 
four electron bunches are injected into the same ESR bucket over a duration of $80$ turns. 
Thanks to the synchrotron radiation, these four newly injected bunches will merge into a single bunch.

At $10~\mathrm{GeV}$, each electron bunch should be replaced every $20$ minutes. The lower two plots in 
Figure~\ref{fig:ss} illustrate the evolution of the proton bunch’s emittance. 
When compared to the blue reference curve without electron replacement, 
the increase in proton emittance per electron 
bunch replacement is less than $1\%$ in the horizontal plane and $5\%$ in the vertical plane. 
Consequently, the increase in proton emittance due to electron replacement remains below $20\%$ 
per hour, which is acceptable in comparison to the intra-beam scattering (IBS) lifetime of $2$ hours.

However, as indicated by the upper two plots in Figure~\ref{fig:ss}, the electron emittance 
experiences significant change. Specifically, in betatron injection with an initial transverse
offset of $7\sigma_x$, the horizontal emittance increases up to $25$ times, 
while in the synchrotron injection scheme with an initial momentum offset of $7\sigma_\delta$, 
the vertical emittance rises up to $6$ times. The substantial increase in horizontal emittance 
conflicts with the small dynamic aperture and leads to particle loss. 
Additionally, the enlarged vertical emittance could enhance intrinsic resonances, 
potentially leading to depolarization. Therefore, it is necessary to develop strategies to 
mitigate this electron emittance blowup.

\section{Two kicker scheme in betatron injection}
\begin{figure}[!htbp]
    \centering
    \includegraphics[width=0.85\columnwidth]{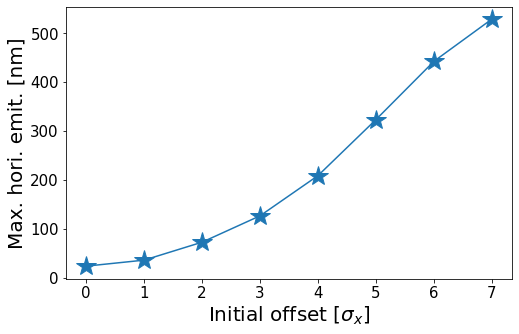}
    \caption{The maximum horizontal emittance versus horizontal injection offset at the IP.}
    \label{fig:maxex}
\end{figure}

The horizontal emittance blowup in betatron injection can be mitigated by reducing the initial offset. 
Figure~\ref{fig:maxex} illustrates the maximum horizontal emittance observed during the strong-strong 
simulation. During the simulation, the initial offset, expressed in terms of $\sigma_x$, 
varies while all other conditions remain consistent with those depicted in Fig.~\ref{fig:ss}. 

\begin{figure}[!htbp]
    \centering
    \includegraphics[width=0.9\columnwidth]{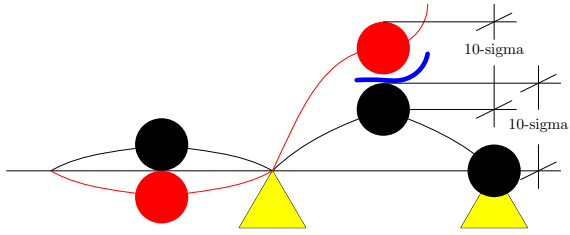}
    \caption{The schematic for the two-kicker scheme in betatron injection. Kickers are 
    shown as yellow triangles. Stored bunches are depicted as black circles, 
    and injected bunches as red circles, each with a radius of $3\sigma$.}
    \label{fig:twoKicker}
\end{figure}

To minimize the initial offset relative to the design orbit while maintaining sufficient separation 
between the injected and stored bunches, the introduction of a second kicker is proposed. 
This concept is illustrated in Fig.~\ref{fig:twoKicker}. As the electron bunch loses polarization, 
the old electron bunch is kicked out. Consequently, when the first bunch is injected, the ESR bucket 
is empty, allowing the bunch to be injected directly on orbit. For subsequent injections, 
the stored bunch is shifted to a position with a negative offset at the injection point, 
while the newly injected 
bunch is positioned with a positive offset. This arrangement ensures that both bunches maintain 
a small offset relative to the design orbit and achieve adequate separation to accommodate 
the septum magnet.

\begin{figure}[!h]
    \centering
    \includegraphics[width=0.85\columnwidth]{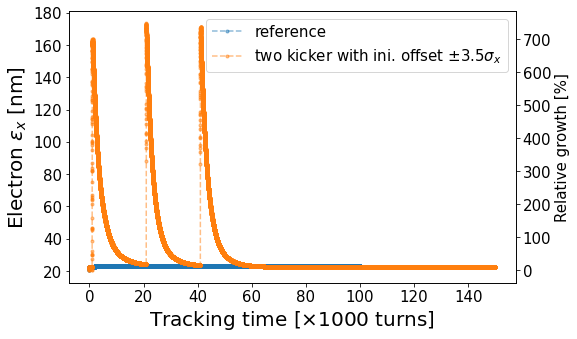}
    \caption{Horizontal emittance evolution with the two kicker scheme. 
    Here strong-strong simulation model is used.}
    \label{fig:twoKickerEmitx}
\end{figure}

Figure~\ref{fig:twoKickerEmitx} displays the simulation results for the two-kicker scheme, 
specifically focusing on the electron's horizontal emittance. 
The duration between two injections is $20,000$ turns, which is sufficient long to merge
the stored and injected beam.
In this setup, 
both the stored and injected bunches are directed to positions of $\pm3.5\sigma_x$ at the 
interaction point (IP). As a result, the maximum horizontal emittance is reduced to seven times 
of the design value. Further examination of the phase space output reveals that only $0.05\%$ 
of macro-particles exceed the $10\sigma_x$ dynamic aperture.

Further reduction of the maximum horizontal emittance is indeed achievable by optimizing the 
phase advance between the IP and the injection point, as well as by utilizing a pulse quadrupole.

\section{Vertical emittance blowup in synchrotron injection}
The electron emittance blowup in synchrotron injection can be attributed to synchro-betatron resonance. 
This phenomenon occurs when an electron bunch is injected with an off-momentum deviation. 
The momentum offset 
translates into a longitudinal offset due to longitudinal oscillation. As the bunch goes through 
the crab cavity, the significant longitudinal offset induces a nonlinear crab kick, 
resulting in a horizontal offset at the IP. This horizontal displacement is 
modulated by the longitudinal motion, which in turn excites higher-order 
synchro-betatron resonances via beam-beam interaction 
\cite{PhysRevAccelBeams.24.041002}.

\begin{figure}[!h]
    \centering
    \includegraphics[width=0.85\columnwidth]{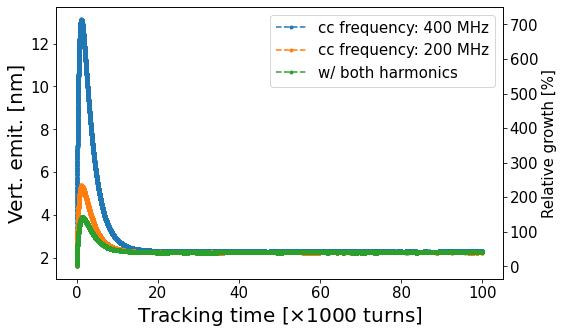}
    \caption{Vertical emittance evolution by weak-strong simulation for difference crab cavity frequencies. The initial momentum offset is $7\sigma_\delta$.}
    \label{fig:harmonicCC}
\end{figure}

Figure~\ref{fig:harmonicCC} illustrates the vertical emittance evolution through a weak-strong simulation. 
The vertical emittance blowup is significantly mitigated by selecting a lower frequency for the crab 
cavity. Utilizing a combination of $200~\mathrm{MHz}$ and $400~\mathrm{MHz}$ frequencies for the crab 
cavity results in a more linear crab cavity kick, which further reduces the emittance blowup. This 
approach clearly demonstrates the emittance blowup is caused by synchro-betatron resonance.

\begin{figure}[!h]
    \centering
    \includegraphics[width=0.85\columnwidth]{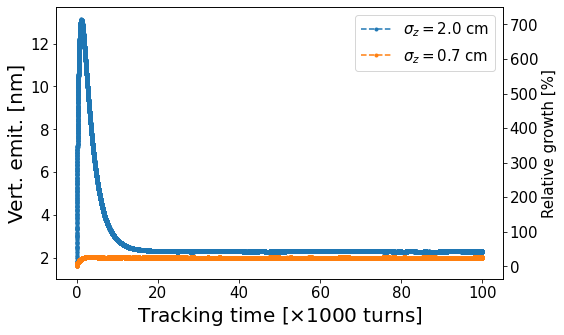}
    \caption{Vertical emittance evolution by weak-strong simulation for difference equilibrium bunch 
    length. The initial momentum offset is $7\sigma_\delta$. The crab cavity frequency 
    is $400~\mathrm{MHz}$.}
    \label{fig:sigz}
\end{figure}

Reducing the longitudinal action can effectively decrease the strength of synchro-betatron resonance. 
Figure~\ref{fig:sigz} shows a comparison of different electron bunch lengths through the weak-strong 
simulation. The term ``bunch length'' here refers to the equilibrium length in the absence of beam-beam 
interaction, which is proportional to the bucket width. When the electron bunch length is reduced 
to $0.7~\mathrm{cm}$, which is the current design value, the vertical emittance blowup is eliminated. 

\section{Conclusion}
One possible method to accumulate a high electron charge is by injecting multiple electron 
bunches into the same ESR bucket. 
This paper investigates betatron and synchrotron injection schemes using both 
weak-strong and strong-strong simulations. A two-kicker scheme is proposed to mitigate 
the horizontal emittance blowup in the betatron injection scheme. 
The vertical emittance blowup in the synchrotron injection scheme can be 
alleviated by reducing the electron bunch length. Given that a bunch length of 
$0.7~\mathrm{cm}$ is included in the latest baseline design of the ESR, synchrotron injection 
emerges as a viable method for merging multiple electron bunches into a single bunch 
within the same bucket.

%
% only for "biblatex"
%
\ifboolexpr{bool{jacowbiblatex}}%
	{\printbibliography}%

@techreport{willeke2021electron,
  title={Electron ion collider conceptual design report 2021},
  author={Willeke, Ferdinand and Beebe-Wang, J},
  year={2021},
  institution={Brookhaven National Lab.(BNL), Upton, NY (United States); Thomas Jefferson National Accelerator Facility (TJNAF), Newport News, VA (United States)}
}

@techreport{lovelace2023damping,
  title={A Damping Ring for the Rapid Cycling Synchrotron},
  author={Lovelace III, Henry and Kewisch, J},
  year={2023},
  institution={Brookhaven National Laboratory (BNL), Upton, NY (United States)}
}

@inproceedings{calvey2021measurements,
  title={Measurements and Simulations of High Charge Beam in the APS Booster},
  author={Calvey, JR and Dooling, JC and Harkay, KC and Wootton, KP and Yao, Chihyuan and others},
  booktitle={12th International Particle Accelerator Conference (IPAC'21), Campinas, SP, Brazil, 24-28 May 2021},
  pages={197--200},
  year={2021},
  organization={JACOW Publishing, Geneva, Switzerland}
}

@article{mori2012design,
  title={Design study of beam injection for SuperKEKB main ring},
  author={Mori, T and Iida, N and Kikuchi, M and Mimashi, T and Sakamoto, Y and Sugimoto, H and Takasaki, S and Tawada, M and others},
  journal={Proceedings of the IPAC},
  volume={12},
  pages={2035--2037},
  year={2012}
}

@article{PhysRevAccelBeams.24.041002,
  title = {Synchrobetatron resonance of crab crossing scheme with large crossing angle and finite bunch length},
  author = {Xu, Derong and Hao, Yue and Luo, Yun and Qiang, Ji},
  journal = {Phys. Rev. Accel. Beams},
  volume = {24},
  issue = {4},
  pages = {041002},
  numpages = {15},
  year = {2021},
  month = {Apr},
  publisher = {American Physical Society},
  doi = {10.1103/PhysRevAccelBeams.24.041002},
  url = {https://link.aps.org/doi/10.1103/PhysRevAccelBeams.24.041002}
}

@inproceedings{qiang:ipac2021-wepab252,
  author       = {J. Qiang and M. Blaskiewicz and Y. Hao and Y. Luo and C. Montag and F.J. Willeke and D. Xu},
  title        = {{Transient Beam-Beam Effect During Electron Bunch Replacement in the EIC}},
  booktitle    = {Proc. IPAC'21},
  pages        = {3228--3231},
  eid          = {WEPAB252},
  venue        = {Campinas, SP, Brazil},
  series       = {International Particle Accelerator Conference},
  number       = {12},
  publisher    = {JACoW Publishing, Geneva, Switzerland},
  month        = {08},
  year         = {2021},
  issn         = {2673-5490},
  isbn         = {978-3-95450-214-1},
  doi          = {10.18429/JACoW-IPAC2021-WEPAB252},
  url          = {https://jacow.org/ipac2021/papers/wepab252.pdf},
  note         = {https://doi.org/10.18429/JACoW-IPAC2021-WEPAB252},
}
	{%
	% "biblatex" is not used, go the "manual" way
	
	%\begin{thebibliography}{99}   % Use for  10-99  references
	
} % end \ifboolexpr
%
% for use as JACoW template the inclusion of the ANNEX parts have been commented out
% to generate the complete documentation please remove the "%" of the next two commands
% 
%%%\newpage

%%%\include{annexes-A4}

\end{document}